\documentclass[
reprint, 
superscriptaddress,
 amsmath,amssymb,
 aps,
]{revtex4-2}

\usepackage[utf8]{inputenc}
\usepackage{graphicx}
\usepackage{dcolumn}
\usepackage{bm}
\usepackage{hyperref}
\usepackage[mathlines]{lineno}
\usepackage{amsmath} 
\usepackage{siunitx}   
\usepackage{multirow}  
\usepackage{physics}
\usepackage{xcolor}
\usepackage{comment} 

\begin{document}

\title{ Slow and Stored Light via Electromagnetically Induced Transparency Using A $\Lambda$-type Superconducting Artificial Atom}

\author{Kai-I Chu}
\affiliation{Molecular Science and Technology Program, Taiwan International Graduate Program, Academia Sinica, Taiwan}
\affiliation{Department of Physics, National Central University, Taoyuan City 32001, Taiwan}

\author{Xiao-Cheng Lu}
\affiliation{Department of Physics, National Tsing Hua University, Hsinchu 30013, Taiwan}
\author{Kuan-Hsun Chiang}
\affiliation{Department of Physics, National Central University, Taoyuan City 32001, Taiwan}
\author{Yen-Hsiang Lin}
\affiliation{Department of Physics, National Tsing Hua University, Hsinchu 30013, Taiwan}
\author{Chii-Dong Chen}
\affiliation{Institute of Physics, Academia Sinica, Taipei 11529, Taiwan}

\author{Ite A. Yu}
\affiliation{Department of Physics, National Tsing Hua University, Hsinchu 30013, Taiwan}
\affiliation{Center for Quantum Science and Technology, National Tsing Hua University, Hsinchu 30013, Taiwan}

\author{Wen-Te Liao}
\affiliation{Department of Physics, National Central University, Taoyuan City 32001, Taiwan}
\affiliation{Quantum Technology Center, National Central University, Taoyuan City 32001, Taiwan}
\affiliation{Physics Division, National Center for Theoretical Sciences, Taipei 10617, Taiwan}
\author{Yung-Fu~Chen}
\email[Correspondence to: ]{yfuchen@ncu.edu.tw}
\affiliation{Department of Physics, National Central University, Taoyuan City 32001, Taiwan}
\affiliation{Quantum Technology Center, National Central University, Taoyuan City 32001, Taiwan}

\begin{abstract}  
Recent progresses in Josephson-junction-based superconducting circuits have propelled quantum information processing forward. However, the lack of a metastable state in most superconducting artificial atoms hinders the development of photonic quantum memory in this platform. Here, we use a single superconducting qubit-resonator system to realize a desired $\Lambda$-type artificial atom, and to demonstrate slow light with a group velocity of 3.6 km/s and the microwave storage with a memory time extending to several hundred nanoseconds via electromagnetically induced transparency. Our results highlight the potential of achieving microwave quantum memory, promising substantial advancements in quantum information processing within superconducting circuits.

\end{abstract}

\date{\today}

\maketitle
\section{Introduction}
Optical quantum memory is a fundamental and crucial ingredient in a quantum network \cite{kimble2008quantum} that can store and retrieve the photonic information on demand. Moreover, it serves as the building block in buffering, synchronizing, and distributing entanglements across various nodes and locations within the network \cite{lvovsky2009optical,tittel2010photon,heshami2016quantum}. To realize the idea of optical quantum memory, the photonic information is typically encoded to the long solid-state coherence through the light-matter interaction. 

In the microwave regime, superconducting circuits have become the prospective platform for quantum computing and information processing owing to the engineerable and strong coupling between the circuits and single photons \cite{wendin2017quantum,arute2019quantum,song2019generation,wang2020controllable,wu2021strong,axline2018demand,kurpiers2018deterministic}. To further build up a quantum network based on superconducting circuits \cite{magnard2020microwave}, the development of microwave quantum memory is an active topic in the field. Over the past decade, numerous schemes for microwave quantum memory have come to fruition, often relying on the resonator's absorption. Besides, the realization of the storage effect is achieved through the manipulation of the coupling between the resonator and the information propagation channel \cite{yin2013catch,wenner2014catching,flurin2015superconducting}, as well as through the inhibition of rephasing processes within resonator arrays \cite{bao2021demand,matanin2023toward}.  

For natural atoms, one of the promising approaches to slow and even store the photons in the atomic media is based on electromagnetically induced transparency (EIT) \cite{phillips2001storage,liu2001observation,kocharovskaya2001stopping,fleischhauer2005electromagnetically}. Notably, the storage efficiencies up to 92\% have been achieved experimentally in cold atom systems \cite{hsiao2018highly}. In EIT media, a weak resonant probe light, which strongly interacts with the atoms, attains transparency due to the presence of a separate, strong coupling field. Simultaneously, the interaction with the coupling field significantly alters the dispersion of the probe light, enabling precise control over the group velocity of the propagating pulse. By adiabatically controlling the coupling field, the photonic information carried by the probe pulse can be effectively stored in and retrieved from the atomic coherence \cite{fleischhauer2000dark}, thus realizing the concept of optical quantum memory \cite{lvovsky2009optical,tittel2010photon,heshami2016quantum,hammerer2010quantum,lukin2003colloquium}.  Efforts to realize EIT in superconducting circuits have encountered limitations due to the lack of suitable metastable states \cite{abdumalikov2010electromagnetically,hoi2011demonstration,anisimov2011objectively}. In the past few years, research progress has been achieved in observing EIT spectroscopy \cite{liu2016method,novikov2016raman,long2018electromagnetically,ann2020tunable,vadiraj2021engineering}. However, their designs and operating methods are not suitable for the development of EIT quantum memory \cite{chu2023three}. A recent slow light experiment was based on the dispersion slope due to Autler-Towns splitting instead of EIT in the qubit metamaterial \cite{brehm2022slowing}. Therefore, investigations into slow and stored light experiments based on EIT have yet to be explored in superconducting circuits.

In this work, we present the experimental realization of EIT slow and stored light phenomena in superconducting waveguide quantum electrodynamics (WQED) architectures, as initially proposed in our recent theoretical investigation \cite{chu2023three}. A distinctive aspect of WQED is its ability to couple artificial atoms with a continuous spectrum of propagating photon modes - a feature suited for addressing the manipulation of cascading photonic information, including light generation, detection, and storage \cite{hoi2011demonstration,roy2017colloquium,sheremet2023waveguide,kannan2023demand,grimsmo2021quantum}. Our approach involves the parametric modulation of the qubit transition frequency, enabling the bridging of the first-order dipole-forbidden sideband transition \cite{blais2007quantum} in a qubit-resonator system \cite{strand2013first,lu2017universal,naik2017random}, and creating an artificial effective $\Lambda$-type atom. By employing this parametric modulation technique, which plays the role of the coupling field utilized in real atom scenarios, the phase response of the probe light is 
significantly modified and becomes steep. A weak probe pulse is decelerated under the continuous parametric modulation to the qubit which results in a reduction in its group velocity to 3.6 km/s. Furthermore, by adiabatically switching off the parametric modulation, we are able to partially store the pulse, characterized by an average photon count below the single photon level, within the resonator. Subsequently, the stored pulse is retrieved back to the transmission line by switching on the modulation again. Besides, by activating the parametric modulation with different magnitudes, the retrieved pulse can be shaped by precisely controlling the group velocity \cite{chen2005manipulating}. This single $\Lambda$-type atom EIT offers the possibility to study the fundamental limits of light-matter interactions and single-atom quantum memory \cite{mucke2010electromagnetically,specht2011single,korber2018decoherence}.

\section{Single Artificial $\Lambda$-Type Atom }
\begin{figure*}[t!]
    \centering
    \includegraphics[width=0.9\textwidth]{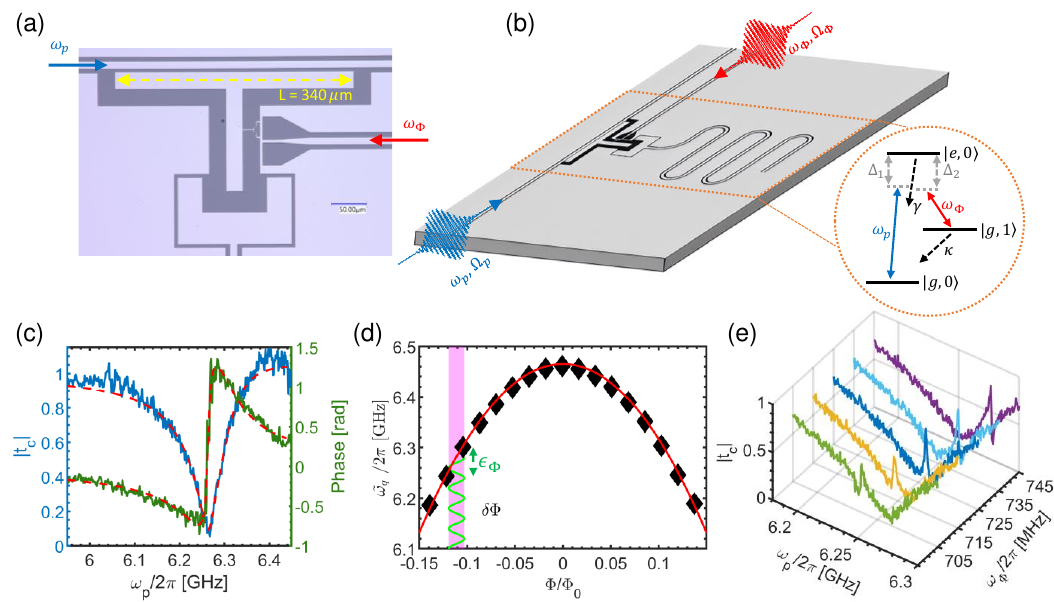}
    \caption{$\Lambda$-type superconducting artificial atom. (a) Optical micrograph of the sample. The length of the qubit (EIT media) is $L=340$ $\mu$m. (b) Schematic of the circuit layout and the corresponding energy-level diagram. The tunable Xmon qubit is capacitively coupled to a transmission line and a detuned $\lambda/4$ resonator. The weak probe light with frequency $\omega_{p}$ is sent to the transmission line and interacts with the qubit. The parametric modulation with frequency $\omega_{\Phi}$ goes through the flux line to modulate the qubit transition frequency. The $\Lambda$-type energy levels are established with the parametric modulation. $\Delta_1=\tilde{\omega}^M_{q}-\omega_p$ and $\Delta_2=\tilde{\omega}^M_{q}-\tilde{\omega}_{r}-\omega_{\Phi}$ denote the field detunings from the motional average qubit frequency $\tilde{\omega}^M_{q}$ induced by modulation. (c) Single-tone spectroscopy. Qubit is biased at $\Phi=-0.11\Phi_0$. The transmission $|t_c|$ (blue) and phase (green) versus $\omega_p$ with $P_p=-141.8$ dBm. The qubit parameters are extracted from the theoretical fits (dashed lines). (d) $\tilde{\omega}_{q}$ as a function of the flux $\Phi$ (black diamonds). The red solid line denotes the theoretical fit. The green sinusoidal wave indicates the flux modulation $\delta\Phi$, which results in the modulation magnitude of the qubit resonance $\epsilon_{\Phi}$. (e) Two-tone spectroscopy as a function of $\omega_{\Phi}$. The splitting is observed with the parametric modulation.}
    \label{fig:setup}
\end{figure*}

Our circuit, illustrated in Fig. \ref{fig:setup}(a) and (b), consists of a tunable superconducting Xmon qubit \cite{barends2013coherent} with the bare transition frequency $\omega_q$ strongly coupled to both a one-dimensional open transmission line and a resonator. The corresponding circuit model and the whole measurement setup are described in Appendix \ref{subsec:Method1}. The microwave probe signal is injected into the transmission line, where it directly interacts with the qubit. The resonator with the bare resonance frequency $\omega_r$, located at a considerable distance from the transmission line, exhibits a small energy loss rate $\kappa$ in comparison to the qubit's total decoherence rate $\gamma$. The resonator serves to store microwave photons. The system is operated in the dispersive regime, where the frequency difference between the qubit and the resonator is much larger than their coupling strength $g_{q,r}$ and loss rates. Fig. \ref{fig:setup}(b) shows the energy level diagram of the $\Lambda$-type atom, which comprises the lowest three energy dressed states $\ket{g,0}$, $\ket{e,0}$, and $\ket{g,1}$, where $g(e)$ denotes the qubit is mainly at the ground (excited) state, and the resonator is at the photon Fock state $0$ or $1$. The probe light directly drives the transition from the ground state $\ket{g,0}$ to the excited state $\ket{e,0}$. Meanwhile, the parametric modulation serves to bridge the otherwise forbidden transition from the excited state $\ket{e,0}$ to the state $\ket{g,1}$, where $\ket{g,1}$ acts as a metastable state due to its longer coherence time compared to the excited state $\ket{e,0}$. As a result, the probe light exhibits transparency through an interference effect induced by the parametric modulation.

Our initial steps involve characterizing the basic properties of the qubit (far detuned from $\omega_r$) by measuring the transmission coefficient $t_c$ of the scattered probe light with single-tone spectroscopy. The bare qubit transition frequency follows $\omega_q(\Phi)=\sqrt{8E_cE_J(\Phi/\Phi_{0})}-E_c$, where $E_c$ denotes the charging energy, $E_J(\Phi/\Phi_{0})$ the flux-dependent Josephson energy, $\Phi$ the applied magnetic flux and $\Phi_{0}$ the flux quantum. Fig. \ref{fig:setup}(c) depicts $|t_c|$ versus the probe frequency $\omega_p$ when the qubit is biased at $\Phi=-0.11\Phi_0$ and Fig. \ref{fig:setup}(d) depicts the dressed qubit transition frequency $\tilde{\omega}_{q}$ varies with $\Phi$. Note that the experiments are done at the weak probe power, such that the qubit is not saturated. The probe light near $\omega_p=\tilde{\omega}_{q}$ is reflected by the qubit as long as the pure dephasing rate of the qubit $\gamma_{\phi}$ is small and the probe Rabi frequency $\Omega_p$ is sufficiently weak compared to the total decoherence rate $\gamma=\gamma_{\phi}+\Gamma/2$, where $\Gamma$ is the relaxation rate of the qubit. Via fitting the transmission profile in Fig. \ref{fig:setup}(c), $\tilde{\omega}_{q}/2\pi=6.282$ GHz, $\Gamma/2\pi=121$ MHz, and $\gamma_{\phi}/2\pi=3$ MHz are extracted \cite{astafiev2010resonance,hoi2011demonstration,probst2015efficient}. By analyzing the saturation of  $t_c$ as a function of the probe power, we extract $\Omega_p$ to calibrate the input probe power $P_p=\hbar\tilde{\omega}_{q}\Omega_p^2/(2\Gamma)$ \cite{lu2021characterizing}. For the spectroscopy experiments, $P_p=-141.8$ dBm, corresponding to $\Omega_p/2\pi=7.8$ MHz, is fixed. At $\Phi=-0.11\Phi_0$, the dressed resonator frequency $\tilde{\omega}_{r}/2\pi=5.532$ GHz, and its energy loss rate $\kappa/2\pi=0.78$ MHz are extracted. Such a large $\kappa$ is mainly caused by the inverse Purcell effect \cite{houck2007generating}. The bare resonance frequency of the resonator $\omega_r/2\pi=5.539$ GHz is observed by a strong probe power $P_p=-96.8$ dBm to saturate the qubit. The coupling $g_{q,r}/2\pi=73.3$ MHz is determined by calculating $g_{q,r}=\sqrt{(\omega_r-\tilde{\omega}_{r})(\omega_q-\omega_r)}$ \cite{blais2021circuit}. The best fit, shown in Figure \ref{fig:setup}(d), gives $E_c/2\pi=290$ MHz, $E_J(\Phi=0)/2\pi=19.6$ GHz, and the junction asymmetry $d=0.32$.

To manipulate EIT with our $\Lambda$-type atom, we send another coherent light to the flux line to induce the flux modulation, i.e., $\Phi(t)=\Phi+\delta\Phi\sin{(\omega_{\Phi}t)}$, where $\delta\Phi$ and $\omega_{\Phi}$
are the modulation magnitude and frequency (see Fig. \ref{fig:setup}(a) and (d)). Modulating the flux $\Phi$ results in $E_J(t)=E_J\cos{(\pi\Phi(t)/\Phi_{0}})\sqrt{1+d^2\tan^2{(\pi\Phi(t)/\Phi_{0})}}$, and therein the qubit transition frequency is being periodically modulated. In Fig. \ref{fig:setup}(e), we present $|t_c|$ versus $\omega_p$ and $\omega_{\Phi}$. The modulation enables the previously forbidden sideband transition of $\ket{e,0}$ to $\ket{g,1}$ with $\omega_{\Phi}\approx\tilde{\omega}_{q}-\tilde{\omega}_{r}$ \cite{chu2023three} and leads to a typical $\Lambda$-type energy level diagram with the parametric modulation Rabi frequency $\Omega_{\Phi}$. The magnitude of $\Omega_{\Phi}$ is mainly determined by \cite{strand2013first,chu2023three}
\begin{equation}
    \label{eq:Parametric rabi}
\Omega_{\Phi}=2g_{q,r}J_{1}(\frac{\epsilon_{\Phi}}{2\omega_{\Phi}}),
\end{equation}
where $J_{1}$ is the Bessel function of the first kind and $\epsilon_{\Phi}$ is the resulting parametric modulation magnitude of the qubit transition frequency through the flux modulation. Under the weak modulation, $\epsilon_{\Phi}$ is proportional to $\delta\Phi$,  which results in $\Omega_{\Phi}\propto\delta\Phi$.

\section{Observation of EIT}
\begin{figure*}[htbp]
    \centering
    \includegraphics[width=0.9\textwidth]{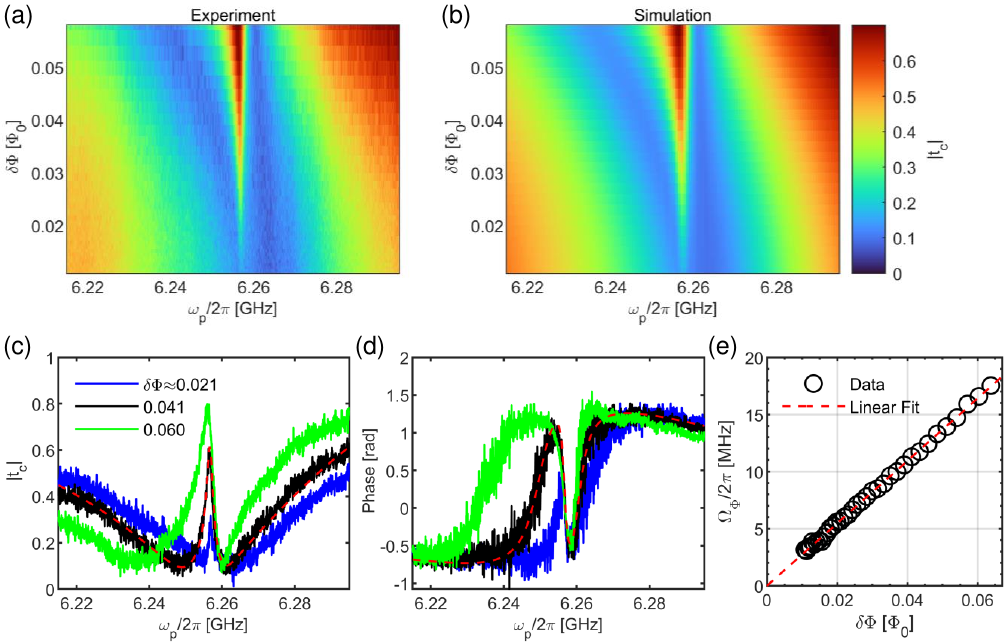}
    \caption{EIT spectroscopy. (a) False-color spectroscopy of $|t_c|$ of the experiment and (b) the simulation as functions of $\delta\Phi$ and $\omega_p$. $\omega_{\Phi}/2\pi=725$ MHz is fixed. (c) Line cuts of the experimental results. (d) The corresponding phase response. The transparency occurs at the two-photon resonance near $\omega^0_p/2\pi=6.2565$ GHz, and the steep change in phase results in the slow light. The fitting results of the experimental data are depicted in red dashed lines. (e) $\Omega_{\Phi}$ as a function of $\delta\Phi$. Black empty circles are the fitting $\Omega_{\Phi}$ of the data in (a). The red dashed line is the linear fit.}
    \label{fig:EIT}
\end{figure*}

Subsequently, we fix $\omega_{\Phi}/2\pi=725$ MHz in the experiments. $|t_c|$ as functions of $\omega_p$ and $\delta\Phi$ are depicted in Fig. \ref{fig:EIT}(a). Line cuts of Fig. \ref{fig:EIT}(a) and the corresponding phase response are shown in Fig. \ref{fig:EIT}(c) and (d). The steady-state response of the probe light in the experiment can be well described via the expectation value $\expval{\sigma}$ with the driven Jaynes-Cummings Hamiltonian (see Fig. \ref{fig:EIT}(b)), written as
\begin{equation}
 \begin{aligned}
    \label{eq:Driving Hamiltonian}
    H(t)= &\omega_{q}(t)\sigma^\dagger\sigma + \omega_{r}a^\dagger a \\&+ g_{q,r}(a^\dagger\sigma + \sigma^\dagger a) + \Omega_{p}\cos{(\omega_{p}t)}(\sigma+\sigma^\dagger),
\end{aligned}
\end{equation}
where $\sigma$ ($\sigma^\dagger$) is the lowering (raising) operator for the qubit and $a$ ($a^\dagger$) is the lowering (raising) operator for the resonator. The simulation is done by solving the master equation by using QuTiP \cite{johansson2012qutip} and is described in Appendix \ref{subsec:Method2}. With the parametric modulation, the probe light is no longer reflected by the qubit; instead, it becomes transparent due to the two-photon resonance \cite{fleischhauer2005electromagnetically}. As $\delta\Phi$ increases, it leads to the expansion of the EIT window and enhances the transparency of the probe light. Besides, due to the nonlinearity of the qubit transition frequency, there is a frequency shift caused by motional averaging effects \cite{li2013motional,beaudoin2012first}. Nevertheless, the frequency at which the two-photon resonance occurs remains well within the transparency window, which enables the study of slow and stored light phenomena. In fact, the optimal operating point necessitates biasing at the qubit linear frequency tuning regime, mitigating the qubit frequency shift arising from the parametric modulation \cite{chu2023three}. However, $\omega_q(\Phi=0)$ is not far greater than $\omega_r$ in this chip. As a result, the following experimental results are conducted with the qubit biased at $\Phi=-0.11\Phi_0$ to have a trade-off between the inverse Purcell effect and the qubit frequency shifts.

To comprehensively investigate the EIT phenomenon in our setup, we employ analytical modeling with the $\Lambda$-type atom interacting with the probe and coupling lights, while considering the frequency shift induced by the parametric modulation on the qubit. Besides, the corresponding frequency shift of the resonator due to the qubit's frequency shift is deemed negligible in our analysis. We define the motional average qubit frequency as $\tilde{\omega}^M_{q}$. The analytical form for the transmission coefficient is written as \cite{gu2017microwave,chu2023three}
 \begin{equation}
    \label{eq:Driven Lambda type}
    t^{a}_{c} = 1 + i\frac{\frac{\Gamma}{2}({\delta-i\frac{\kappa}{2})}}{(\delta-i\frac{\kappa}{2})(\delta+\Delta_2-i\gamma)-(\frac{\Omega_{\Phi}^2}{4})},
\end{equation}
where $\Delta_1=\tilde{\omega}^M_{q}-\omega_p$, $\Delta_2=\tilde{\omega}^M_{q}-\tilde{\omega}_{r}-\omega_{\Phi}$, and $\delta=\Delta_1-\Delta_2$. The fitting parameters are $\tilde{\omega}^M_{q}$ and $\Omega_{\Phi}$. Other parameters are extracted from single-tone spectroscopy. The fitting matches the data shown in Fig. \ref{fig:EIT}(c) and (d), verifying that our system behaves as a $\Lambda$-type atom. The probe light at a frequency near $\omega_p\approx\tilde{\omega}^M_{q}\equiv\omega^0_p=2\pi\times6.2565$ GHz becomes transparent. Due to the condition of $\gamma>>\kappa/2$, the optical response is fully governed by EIT as long as the parametric modulation satisfies the situation $\Omega_{\Phi}<\gamma$. The pulsed experimental data shown later also verify the observation of EIT. Fig. \ref{fig:EIT}(e) shows the extracted $\Omega_{\Phi}$, from the fitting results in Fig. \ref{fig:EIT}(a), as a function of $\delta\Phi$. As expected, $\Omega_{\Phi}$ is linearly increased with $\delta\Phi$.

\section{Characterization of Slow Light}
\begin{figure*}[t!]
    \centering
    \includegraphics[width=0.9\textwidth]{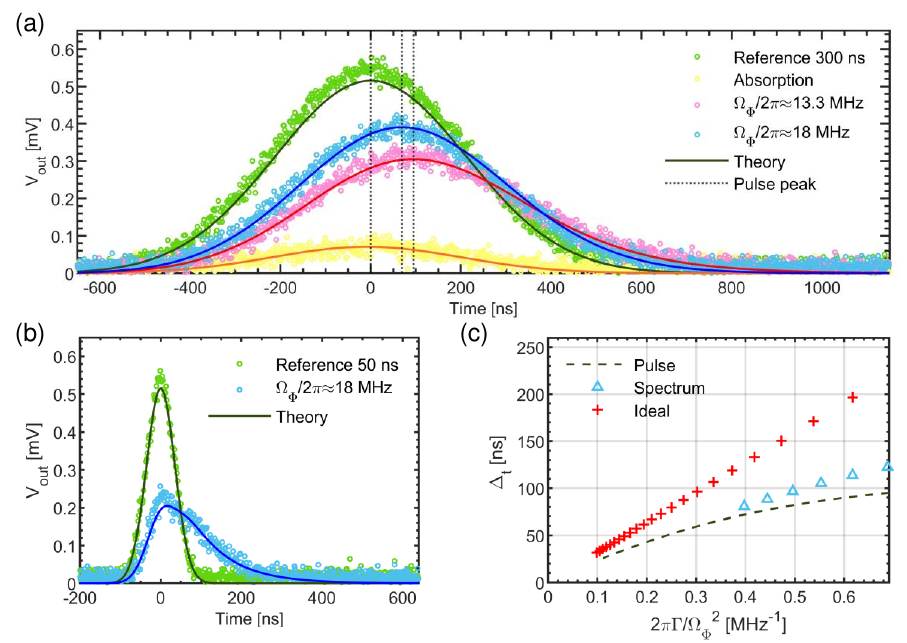}
    \caption{Demonstration of the slow light. (a) Probe pulse envelopes with different continuous $\Omega_{\Phi}$. See the text.  The pulse with the parametric modulation becomes slower than the reference. The solid lines with different colors are the theoretical predictions of the slow light. The black dotted lines, defined as the pulse peaks, are the times when the maximum amplitudes of the theoretical predictions occur. The corresponding delay times are $\Delta_t= 95$ and $69$ ns, respectively. (b) The shorter pulse duration of probe pulse envelopes with continuous $\Omega_{\Phi}$. It will be used for the later stored light experiments, shown in Fig. \ref{fig:Trappedlight}. (c) $\Delta_t$ versus $\Gamma/\Omega_{\Phi}^2$. The dark green dashed line is extracted from the theoretical prediction of the pulse experiments in (a). The light blue empty triangles are analyzed from the spectroscopy data and Eq. \ref{eq: phase derivation}. The red plus-sign markers are the ideal case, which gives $D=2$.}
    \label{fig:slowlight}
\end{figure*}

The rapid change of the phase at the two-photon resonance with the parametric modulation indicates the reduction of the group velocity. In order to investigate the modified group velocity, The delay time of the probe pulse is experimentally measured by transmitting a Gaussian probe pulse through the transmission line. The pulse envelope is expressed as $\Omega^s_p\exp(-t^2/{\tau_d}^2)$ with $\Omega^s_p/2\pi=7$ MHz and $\tau_d=300$ ns. The corresponding probe power is $P^s_p=5.27\times10^{-18}$ $\si{\watt}$ and the carrier frequency of the pulse is fixed at $\omega^0_p/2\pi$. The phenomenon of the slow light is depicted in Fig. \ref{fig:slowlight}(a). The Gaussian pulse (green empty circles), serving as the reference pulse, is acquired by adjusting the qubit transition frequency away from $\omega^0_p/2\pi$. The pulse (yellow empty circles) represents the interference pattern resulting from the probe pulse and the field emitted by the qubit under a bias of $\Phi=-0.11\Phi_0$. In contrast to the reference pulse, the amplitude of the yellow pulse decreases. The pulses (blue and pink empty circles) are obtained under continuous parametric modulation but $\Omega_{\Phi}/2\pi=18$ MHz and $13.3$ MHz, respectively. Remarkably, both blue and pink pulses exhibit a delay compared to the reference pulse, indicative of the phenomenon of slow light. Moreover, these pulses manifest transparency effects in contrast to the case of absorption, signifying the presence of EIT. It is worth noting that achieving perfect transparency is limited by the finite coherence of the metastable state ($\kappa\neq0$); however, the transparency can be enhanced by increasing $\Omega_{\Phi}$. Therefore, the amplitude of the pulse with higher $\Omega_{\Phi}$ surpasses that of lower $\Omega_{\Phi}$. The numerical calculations of the slow light with different $\Omega_{\Phi}$, depicted as the solid lines, match well with the experimental results. The pulsed delay time $\Delta_t$ is therein defined as the time difference of the maximum amplitude of the theoretical output pulse with the parametric modulation to the center of the reference.

It is known that with the perfect metastable state ($\kappa=0$), the relation between $\Delta_t$ and $\Omega_{\Phi}$ in an ensemble of identical atoms is \cite{fleischhauer2005electromagnetically}
\begin{equation}
    \label{eq:delay time}
\Delta_t = \frac{D\Gamma}{\Omega_{\Phi}^2},
\end{equation}
where $D$ is the optical depth. We further use the relation of Eq. \ref{eq:delay time} to define the effective optical depth of our single atom. Fig. \ref{fig:slowlight}(c) depicts $\Delta_t$ as a function of $\Gamma/\Omega_{\Phi}^2$. $\Delta_t$ is also estimated by measuring the derivative of the phase $\theta$, obtained from the fittings, at $\omega^0_p$ (see Fig. \ref{fig:EIT}(d)), written as  \cite{novikov2016raman,long2018electromagnetically}
\begin{equation}
    \label{eq: phase derivation}
    \Delta_t = \left. -\frac{d\theta}{d\omega_p} \right|_{\omega_p=\omega^0_p}.
\end{equation}
The estimations of $\Delta_t$ from the results of the pulsed and spectrum experiments are consistent. However, when taking into account the non-zero decoherence rate of the metastable state, the maximum achievable $\Delta_t$ becomes limited even if $\Omega_{\Phi}$ is weak \cite{fleischhauer2005electromagnetically}. Besides, the ideal case of $\Delta_t$ in our system, shown as the red plus-sign markers, is obtained via the phase derivation of the two-photon resonance of the Eq. \ref{eq:Driven Lambda type} by assuming $\Delta_1$, $\Delta_2$, and $\kappa=0$. Compared the ideal case to Eq. \ref{eq:delay time}, $D=2$ for a single atom is observed.

Specifically, in the pulsed experiment, the maximum $\Delta_t$ is observed to be approximately $95$ ns at $\Omega_{\Phi}/2\pi= 13.3$ MHz. We consider the length of the medium, i.e., the interaction length of the qubit $L=340$ $\mu$m (see Fig. \ref{fig:setup}(a)) to estimate the group velocity $v_g = L/\Delta_t$. The minimum group velocity of the pulse is reduced to approximately 3.6 km/s. Note that $v_g$ can be further reduced as we decrease $\Omega_{\Phi}$. However, $\Omega_{p}$ should also be reduced to satisfy the EIT condition, i.e., $\Omega_{p}<\Omega_{\Phi}$ .

\section{Single Atom EIT Stored Light}

\begin{figure*}[tbp]
    \centering
    \includegraphics[width=0.9\textwidth]{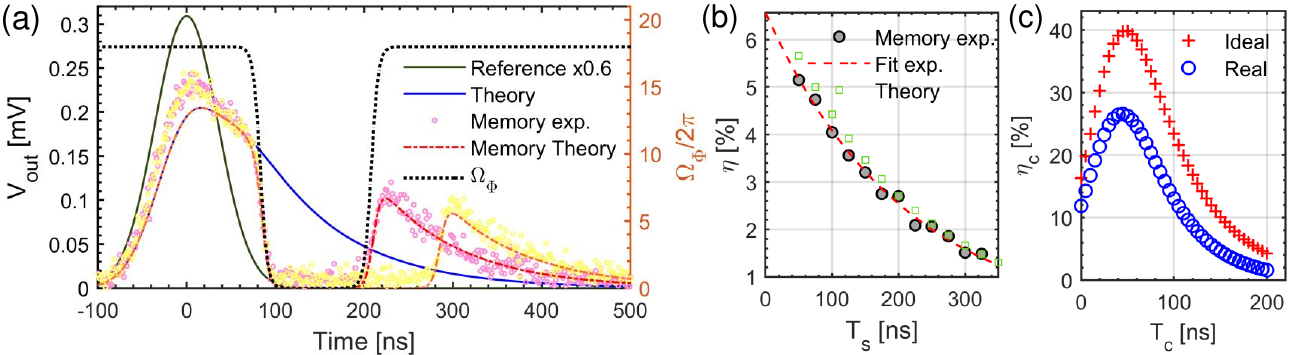}
    \caption{EIT microwave memory with a single atom. (a) Stored light of the pulse. The dark green and blue solid lines represent the reference and slow light. The parametric modulation, shown in the black dotted line, is adiabatically turned off after the Gaussian signal arrives and turned on after $T_s$. The pink (yellow) empty circles after $T_s=125$ $(200)$ ns show the stored-and-retrieval pulse. The theoretical prediction of stored light is depicted with a red (orange) dash-dotted line. (b) Storage efficiency $\eta$ (gray solid circles) varies with $T_s$. The theoretical prediction of $\eta$ is shown in green empty squares. The red dashed line is the decay fit of the experimental data. (c) Captured efficiency $\eta_c$ varies with the modulation turn-off time $T_c$. The red plus-sign markers denote the ideal case, and the blue empty circles denote the real case.}
    \label{fig:Trappedlight}
\end{figure*}

For the stored light experiment, we choose a shorter Gaussian probe pulse compared to the previous experiment. The corresponding full width at half maximum (FWHM) of the pulsed spectral bandwidth is larger than the FWHM of the EIT window $\gamma_{EIT}$ with $D=2$, written as $\gamma_{EIT}\approx\sqrt{\ln{2}}\Omega_{\Phi}^2/(2\sqrt{\Gamma\gamma})$ \cite{hsiao2018highly}. Therefore, the output pulsed shape is distorted, but it allows us to determine the turn-off time of the parametric modulation for the case of the small $D$. A $\tau_d=50$ ns Gaussian probe pulse with the maximum $\Omega_{\Phi}/2\pi=18$ MHz is used, and its slow light is shown in Fig. \ref{fig:slowlight}(b). The spectral FWHM of the probe pulse is about 10.6 MHz and $\gamma_{EIT}/2\pi=1.54$ MHz. The input average photon number of the pulse is $\expval{N_R}= \int [P^s_p(t)/(\hbar\omega^0_p)]dt=0.08$. Fig. \ref{fig:Trappedlight}(a) shows the stored light results. By adiabatically turning off the parametric modulation at the end of the probe pulse, a portion of the pulse is stored within the resonator. Subsequently, through adiabatically turning on the parametric modulation after the storage time $T_s$, the pulse is retrieved back to the transmission line while preserving its original shape. For longer $T_s$, the retrieved pulse becomes weaker. The storage efficiency $\eta$, defined as the energy ratio of the retrieved pulse to the reference pulse, is plotted as a function of $T_s$ (see Fig. \ref{fig:Trappedlight}(b)). The maximum $\eta$ is up to 5\%. The fitting decay rate is about $0.76$ MHz, which is close to the coherence of the metastable state $\kappa$ extracted from the spectroscopy. We highlight that the photon retrieved in our system has the potential to serve as an on-demand single photon source as it relies on the utilization of single atom EIT.

To enhance memory efficiency, we could replace the open transmission line with a semi-infinite transmission line to prevent the retrieved pulse from being emitted in the undesired direction. Enhancing the coherence of the metastable state by eliminating the inverse Purcell effect could also improve memory efficiency. Fig. \ref{fig:Trappedlight}(c) depicts the theoretical study of the captured efficiency $\eta_c$ of the resonator, defined as the ratio of the photon number in the resonator $\expval{a^\dagger a}$ to $\expval{N_R}$, as a function of the time when turning off the modulation $T_c$. $\expval{a^\dagger a}$ is measured after the modulation is almost turned off. We observe that $\eta_c$ in an ideal situation can be up to 40 \% (considering the system with $\kappa=\gamma_{\phi}=0$) and $\eta_c$ can be up to 25 \% in a real situation.

\begin{figure}[tbp]
    \centering
    \includegraphics[width=0.47\textwidth]{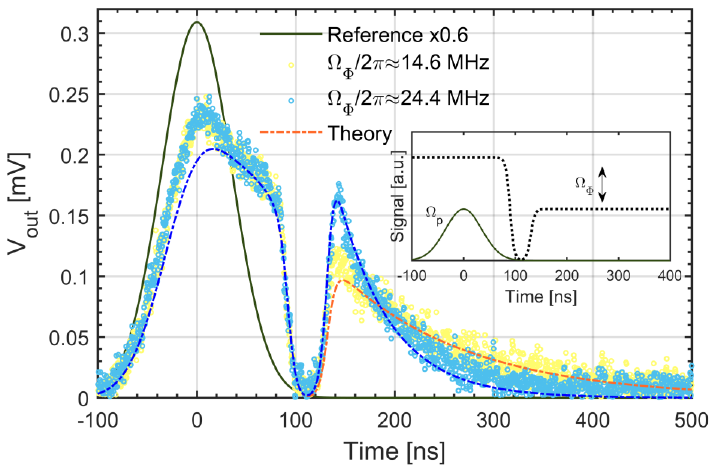}
    \caption{Pulse shaping phenomenon. The retrieved pulse shape after $T_s=40$ is manipulated with $\Omega_{\Phi}/2\pi=14.6$ and $24.4$ MHz, respectively. The inset shows the pulse sequence. }
    \label{fig:shapelight}
\end{figure}
Due to the controllable group velocity of the probe pulse \cite{chen2005manipulating,liu2001observation}, we can manipulate the shape of the retrieved probe pulse by the intensity of the parametric modulation in the retrieved process. The retrieved pulses via different turned-on $\Omega_{\Phi}$ are depicted in Fig. \ref{fig:shapelight}. With the stronger $\Omega_{\Phi}$, the shape of the pulse becomes sharper. Besides, with the weaker $\Omega_{\Phi}$, the shape of the pulse becomes flatter due to the reduction of the group velocity.

\section{Conclusion}
This study presents the experimental demonstration of slow and stored light phenomena using a $\Lambda$-type superconducting artificial atom. The $\Lambda$-type atom is realized through a detuned coupled qubit-resonator system. By introducing parametric modulation, the weak two-photon resonant probe light exhibits transparency, accompanied by a steep dispersion profile. A Gaussian probe pulse under continuous parametric modulation shows a significant slowing down effect, resulting in a delay time of 95 ns and a reduced group velocity of 3.6 km/s. Furthermore, by adiabatically switching off the parametric modulation, a portion of the Gaussian pulse is successfully stored in the resonator. After a designated storage time, the pulse, with its shape controllable, is retrieved back to the transmission line by switching on the parametric modulation with varying magnitudes. This simple yet versatile device provides a platform for investigating different optical quantum memory schemes \cite{reim2010towards,saglamyurek2018coherent} based on superconducting circuits. The outcomes of our work open up new avenues for expanding quantum information processing in the superconducting circuit community.

Future progress in this scheme will focus on the following directions. Firstly, in our prototype, the slow light cannot be separated from the reference pulse due to the small optical depth. Enlarging the optical depth by increasing the interaction length for the qubit \cite{vadiraj2021engineering} or increasing the number of atoms might offer the possibility of completely separating the slow light from the reference pulse, thereby achieving high memory efficiency. Secondly, the current scheme cannot stay the $\Lambda$-type structure at higher probe powers. Via replacing the resonator with another qubit will be explored to enhance the nonlinearity of the dressed states to maintain the $\Lambda$-type structure even at higher probe powers. Consequently, we can make sure there is always a photon interacting with our single atom, which opens up the possibility of generating high-efficiency on-demand single photon sources through EIT.

\begin{acknowledgments}
WTL acknowledges support from Center for Quantum Technology, Hsinchu 30013, Taiwan. The work was supported by the National Science and Technology Council in Taiwan through Grants No. NSTC-110-2112-M-008-024, NSTC-111-2112-M-008-022, NSTC-110-2112-M- 008-027-MY3, NSTC-111-2923-M- 008-004-MY3 and NSTC-111-2639-M-007-001-ASP.
\end{acknowledgments}

\appendix
\section{Experimental Setup}\label{subsec:Method1}
\begin{figure}[htbp]
    \centering
    \includegraphics[width=0.35\textwidth]{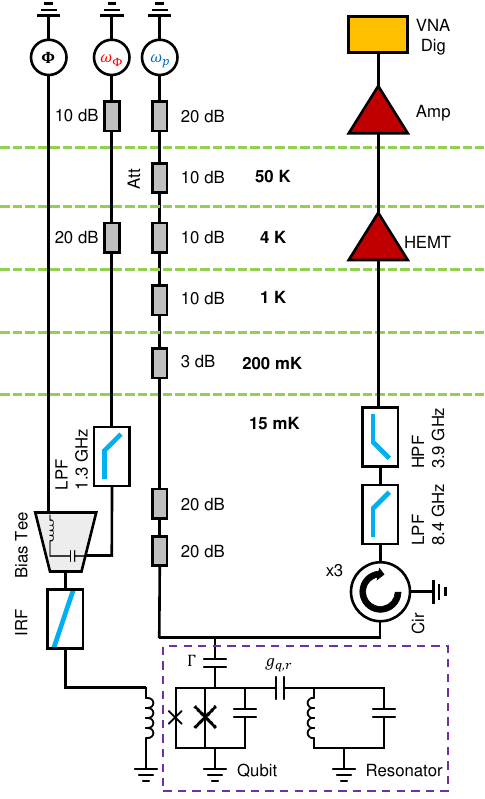}
    \caption{Measurement Setup and schematic diagram of the circuit. Att, Cir, LPF, HPF, IRF, HEMT, Amp, Dig, and VNA denote attenuator, circulator, low-pass filter, high-pass filter, IR filter, high electron mobility transistor amplifier, room-temperature amplifier, digitizer, and vector network analyzer, respectively. See the text. The Circuit model of the chip is denoted by the dashed-line rectangle. The direct coupling between the resonator and the transmission line is neglected due to the significant distance between them \cite{chu2023three}. }
    \label{fig:measurement setup}
\end{figure}

The diagram of the measurement setup is depicted in Fig.
 \ref{fig:measurement setup}. The chip is cooled down through the dilution refrigerator at a temperature of about 20 mK. A vector network analyzer is used to measure the transmission coefficient $|t_{c}|$ and the phase response of the probe light. A digitizer with high sampling rates is used to digitize a short pulsed probe signal. The probe light with frequency $\omega_p$ is sent into the transmission line to capacitively interact with the atom with an attenuation of about 93 dB. The parametric modulation with frequency $\omega_\Phi$ with an attenuation of about 30 dB passing through a low-pass filter (1.3 GHz) is combined with the DC current source $\Phi$ through a bias tee and goes through an IR filter to the qubit's flux line. The transmitted signal is passed through the cryogenic circulators, high-pass filter (8.4GHz), and low-pass filter (3.9 GHz) and is amplified by a high-electron-mobility-transistor amplifier and a few room-temperature amplifiers before reaching the vector network analyzer. The total gain of the output amplification chain is about 92 dB. Besides, the transmitted pulsed signal is demodulated through the IQ mixer with a local oscillator frequency at the pulsed carrier frequency before reaching the digitizer. 

\section{Theoretical Derivation}
\label{subsec:Method2}
In our numerical simulation, we apply the rotating frame at $\omega_{p}$ to transform Eq. \ref{eq:Driving Hamiltonian} and neglect the fast rotating terms to simplify the calculation. The Hamiltonian after the unitary transformation as $H'$ is written as
 \begin{equation}
 \begin{aligned}
    \label{eq:Driving Hamiltonian2}
    H'=&(\omega_{q}(t)-\omega_{p})\sigma^\dagger\sigma + (\omega_{r}-\omega_{p})a^\dagger a \\&+ g_{q,r}(a^\dagger\sigma + \sigma^\dagger a)+\frac{\Omega_{p}}{2}(\sigma^\dagger+\sigma).
\end{aligned}
\end{equation}

The dynamics of the density matrix $\rho(t)$ of the system is calculated by solving the Lindblad master equation:
\begin{equation}
\begin{aligned}
    \label{eq:Master eq}
    \dot{\rho}=&-i[H',\rho]\\&+\frac{\Gamma}{2}\mathcal{D}[\sigma]\rho+\frac{\kappa}{2}\mathcal{D}[a]\rho+\gamma_{\phi}\mathcal{D}[\sigma^\dagger\sigma]\rho,
\end{aligned}
\end{equation}
where $\mathcal{D}[\mathcal{O}]\rho=2\mathcal{O}\rho\mathcal{O}^\dagger-\rho\mathcal{O}^\dagger\mathcal{O}-\mathcal{O}^\dagger\mathcal{O}\rho$.
The transmission coefficient of the steady-state optical response of the scattered probe light is governed by $ t_{c} = 1 + i\Gamma\langle{\sigma}\rangle/\Omega_{p}$.

The pulsed dynamics oscillates near the frequency $\omega_{\Phi}$ when solving the master equation with Eq. \ref{eq:Driving Hamiltonian2}. However, the envelope of the pulse is measured by the homodyne demodulation, where the fast oscillation will be filtered out due to the limited digitizer bandwidth. To better compare the simulation of the dynamics of the envelope of the probe pulse to the experimental results, shown in Fig. \ref{fig:slowlight}-\ref{fig:shapelight}, we derive the time-independent Hamiltonian for the simulation. We first assume that there is no frequency shift induced by the parametric modulation to simplify the discussion, i.e., $ \omega_{q}(t) \approx \omega_{q} + \epsilon_{\Phi}\sin{(\omega_{\Phi}t)}/2$. We apply the unitary transform $U^1_{rot}$ in the rotating frame of instantaneous oscillation frequency to Eq. \ref{eq:Driving Hamiltonian2}, written as
\begin{equation}
\begin{aligned}
    U^1_{rot} = \exp\bigg(&i((\omega_q-\omega_p)t - \frac{\epsilon_{\Phi}}{2\omega_{\Phi}}\cos{(\omega_{\Phi}t)})\sigma^\dagger\sigma\\
    &+ i(\omega_r-\omega_p)t a^\dagger a \bigg).
\end{aligned}
\end{equation}
We consider that $\epsilon_{\Phi}\ll2\omega_{\Phi}$, which results in $\Omega_p$ is not affected by the parametric modulation \cite{chu2023three}. The resulting Hamiltonian is written as \begin{equation}
    \label{eq: 1Hamiltonian}
    \begin{aligned}
    H' =&g_{q,r}\sum_{n=-\infty}^{\infty}J_{n}(\frac{\epsilon_{\Phi}}{2\omega_{\Phi}})[i^n e^{i(n\omega_{\Phi}-\Delta_{q,r})t}a^\dagger\sigma + \textrm{H.c.}]\\&+\frac{\Omega_p}{2}[e^{i(-\omega_q+\omega_p)t}\sigma+\textrm{H.c.}],
    \end{aligned}
    \end{equation}
where $\Delta_{q,r}=\omega_q-\omega_r$ and $J_n$ is the Bessel function of the first kind. The time-dependent terms in Eq. \ref{eq: 1Hamiltonian} are removed by applying another unitary transform \cite{kervinen2019landau}, given as 
\begin{equation}
    \label{eq:Unitary operator2}
    U^2_{rot} = \exp(-i(\omega_q-\omega_p)t\sigma^\dagger\sigma-i(\omega_r-\omega_p+\omega_{\Phi})ta^\dagger a).
\end{equation}
The final Hamiltonian, after neglecting fast rotating terms, is written as
\begin{equation}
\begin{aligned}
\label{eq:Final Hamiltonian}
H'=&(\omega_q-\omega_p)\sigma^\dagger\sigma+(\omega_r-\omega_p+\omega_{\Phi})a^\dagger a\\&+\frac{i\Omega_{\Phi}}{2}(a^\dagger\sigma-a\sigma^\dagger)+\frac{\Omega_p}{2}(\sigma+\sigma^\dagger),
\end{aligned}
\end{equation}
where $\Omega_{\Phi}$ is given in Eq. \ref{eq:Parametric rabi}. To compare with the experimental results, we should consider the frequency shift induced by the parametric modulation. Note that the frequency shift and $\Omega_{\Phi}$ are proportional to ${\delta\Phi}^2$ and ${\delta\Phi}$ (the lowest order in $\delta\Phi$) \cite{beaudoin2012first}. We therein replace $\omega_q$ and $\Omega_{\Phi}$ with $\omega_q-C_{0}{\delta\Phi}^2$ and $C_{1}{\delta\Phi}$, where $C_{0}$ and $C_{1}$ are constants, which can be extracted by analyzing the result in Fig. \ref{fig:EIT}(b). Finally, the magnitude of the output coherent pulse response is calculated with the input-output theory, written as $|\alpha_{out}|=\left|\alpha_{in}+i\sqrt{\Gamma/2}\langle{\sigma}\rangle\right|$, where $\alpha_{in}=\Omega_p/\sqrt{2\Gamma}$.


\begin{thebibliography}{0}
\bibitem{kimble2008quantum}Kimble, H. The quantum internet. {\em Nature}. \textbf{453}, 1023-1030 (2008)
\bibitem{lvovsky2009optical}Lvovsky, A., Sanders, B. \& Tittel, W. Optical quantum memory. {\em Nature Photonics}. \textbf{3}, 706-714 (2009)
\bibitem{tittel2010photon}
Tittel, W., Afzelius, M., Chaneliere, T., Cone, R., Kröll, S., Moiseev, S., \& Sellars, M. (2010).Photon-echo quantum memory in solid state systems.
{\em Laser \& Photonics Reviews}, \textbf{4}, 244-267.
\bibitem{heshami2016quantum}Heshami, K., England, D., Humphreys, P., Bustard, P., Acosta, V., Nunn, J. \& Sussman, B. Quantum memories: emerging applications and recent advances. {\em Journal Of Modern Optics}. \textbf{63}, 2005-2028 (2016)
\bibitem{wendin2017quantum}Wendin, G. Quantum information processing with superconducting circuits: a review. {\em Reports On Progress In Physics}. \textbf{80}, 106001 (2017)
\bibitem{arute2019quantum}Arute, F., Arya, K., Babbush, R., Bacon, D., Bardin, J., Barends, R., Biswas, R., Boixo, S., Brandao, F., Buell, D. \& Others Quantum supremacy using a programmable superconducting processor. {\em Nature}. \textbf{574}, 505-510 (2019)
\bibitem{song2019generation}Song, C., Xu, K., Li, H., Zhang, Y., Zhang, X., Liu, W., Guo, Q., Wang, Z., Ren, W., Hao, J. \& Others Generation of multicomponent atomic Schrödinger cat states of up to 20 qubits. {\em Science}. \textbf{365}, 574-577 (2019)
\bibitem{wang2020controllable}Wang, Z., Li, H., Feng, W., Song, X., Song, C., Liu, W., Guo, Q., Zhang, X., Dong, H., Zheng, D. \& Others Controllable switching between superradiant and subradiant states in a 10-qubit superconducting circuit. {\em Physical Review Letters}. \textbf{124}, 013601 (2020)
\bibitem{wu2021strong}Wu, Y., Bao, W., Cao, S., Chen, F., Chen, M., Chen, X., Chung, T., Deng, H., Du, Y., Fan, D. \& Others Strong quantum computational advantage using a superconducting quantum processor. {\em Physical Review Letters}. \textbf{127}, 180501 (2021)
\bibitem{axline2018demand}Axline, C., Burkhart, L., Pfaff, W., Zhang, M., Chou, K., Campagne-Ibarcq, P., Reinhold, P., Frunzio, L., Girvin, S., Jiang, L. \& Others On-demand quantum state transfer and entanglement between remote microwave cavity memories. {\em Nature Physics}. \textbf{14}, 705-710 (2018)
\bibitem{kurpiers2018deterministic}Kurpiers, P., Magnard, P., Walter, T., Royer, B., Pechal, M., Heinsoo, J., Salathé, Y., Akin, A., Storz, S., Besse, J. \& Others Deterministic quantum state transfer and remote entanglement using microwave photons. {\em Nature}. \textbf{558}, 264-267 (2018)
\bibitem{magnard2020microwave}Magnard, P., Storz, S., Kurpiers, P., Schär, J., Marxer, F., Lütolf, J., Walter, T., Besse, J., Gabureac, M., Reuer, K. \& Others Microwave quantum link between superconducting circuits housed in spatially separated cryogenic systems. {\em Physical Review Letters}. \textbf{125}, 260502 (2020)
\bibitem{yin2013catch}Yin, Y., Chen, Y., Sank, D., O’Malley, P., White, T., Barends, R., Kelly, J., Lucero, E., Mariantoni, M., Megrant, A. \& Others Catch and release of microwave photon states. {\em Physical Review Letters}. \textbf{110}, 107001 (2013)
\bibitem{wenner2014catching}Wenner, J., Yin, Y., Chen, Y., Barends, R., Chiaro, B., Jeffrey, E., Kelly, J., Megrant, A., Mutus, J., Neill, C. \& Others Catching time-reversed microwave coherent state photons with 99.4\% absorption efficiency. {\em Physical Review Letters}. \textbf{112}, 210501 (2014)
\bibitem{flurin2015superconducting}Flurin, E., Roch, N., Pillet, J., Mallet, F. \& Huard, B. Superconducting quantum node for entanglement and storage of microwave radiation. {\em Physical Review Letters}. \textbf{114}, 090503 (2015)
\bibitem{bao2021demand}Bao, Z., Wang, Z., Wu, Y., Li, Y., Ma, C., Song, Y., Zhang, H. \& Duan, L. On-demand storage and retrieval of microwave photons using a superconducting multiresonator quantum memory. {\em Physical Review Letters}. \textbf{127}, 010503 (2021)
\bibitem{matanin2023toward}Matanin, A., Gerasimov, K., Moiseev, E., Smirnov, N., Ivanov, A., Malevannaya, E., Polozov, V., Zikiy, E., Samoilov, A., Rodionov, I. \& Others Toward Highly Efficient Multimode Superconducting Quantum Memory. {\em Physical Review Applied}. \textbf{19}, 034011 (2023)
\bibitem{fleischhauer2005electromagnetically}Fleischhauer, M., Imamoglu, A. \& Marangos, J. Electromagnetically induced transparency: Optics in coherent media. {\em Reviews Of Modern Physics}. \textbf{77}, 633 (2005)
\bibitem{phillips2001storage}Phillips, D., Fleischhauer, A., Mair, A., Walsworth, R. \& Lukin, M. Storage of light in atomic vapor. {\em Physical Review Letters}. \textbf{86}, 783 (2001)
\bibitem{liu2001observation}Liu, C., Dutton, Z., Behroozi, C. \& Hau, L. Observation of coherent optical information storage in an atomic medium using halted light pulses. {\em Nature}. \textbf{409}, 490-493 (2001)
\bibitem{kocharovskaya2001stopping}Kocharovskaya, O., Rostovtsev, Y. \& Scully, M. Stopping light via hot atoms. {\em Physical Review Letters}. \textbf{86}, 628 (2001)
\bibitem{hsiao2018highly}Hsiao, Y., Tsai, P., Chen, H., Lin, S., Hung, C., Lee, C., Chen, Y., Chen, Y., Ite, A. \& Chen, Y. Highly efficient coherent optical memory based on electromagnetically induced transparency. {\em Physical Review Letters}. \textbf{120}, 183602 (2018)
\bibitem{fleischhauer2000dark}Fleischhauer, M. \& Lukin, M. Dark-state polaritons in electromagnetically induced transparency. {\em Physical Review Letters}. \textbf{84}, 5094 (2000)
\bibitem{lukin2003colloquium}Lukin, M. Colloquium: Trapping and manipulating photon states in atomic ensembles. {\em Reviews Of Modern Physics}. \textbf{75}, 457 (2003)
\bibitem{hammerer2010quantum}Hammerer, K., Sørensen, A. \& Polzik, E. Quantum interface between light and atomic ensembles. {\em Reviews Of Modern Physics}. \textbf{82}, 1041 (2010)
\bibitem{abdumalikov2010electromagnetically}Abdumalikov Jr, A., Astafiev, O., Zagoskin, A., Pashkin, Y., Nakamura, Y. \& Tsai, J. Electromagnetically induced transparency on a single artificial atom. {\em Physical Review Letters}. \textbf{104}, 193601 (2010)



\bibitem{hoi2011demonstration}Hoi, I., Wilson, C., Johansson, G., Palomaki, T., Peropadre, B. \& Delsing, P. Demonstration of a single-photon router in the microwave regime. {\em Physical Review Letters}. \textbf{107}, 073601 (2011)
\bibitem{anisimov2011objectively}Anisimov, P., Dowling, J. \& Sanders, B. Objectively discerning Autler-Townes splitting from electromagnetically induced transparency. {\em Physical Review Letters}. \textbf{107}, 163604 (2011)
\bibitem{liu2016method}Liu, Q., Li, T., Luo, X., Zhao, H., Xiong, W., Zhang, Y., Chen, Z., Liu, J., Chen, W., Nori, F. \& Others Method for identifying electromagnetically induced transparency in a tunable circuit quantum electrodynamics system. {\em Physical Review A}. \textbf{93}, 053838 (2016)
\bibitem{novikov2016raman}Novikov, S., Sweeney, T., Robinson, J., Premaratne, S., Suri, B., Wellstood, F. \& Palmer, B. Raman coherence in a circuit quantum electrodynamics lambda system. {\em Nature Physics}. \textbf{12}, 75-79 (2016)
\bibitem{long2018electromagnetically}Long, J., Ku, H., Wu, X., Gu, X., Lake, R., Bal, M., Liu, Y. \& Pappas, D. Electromagnetically induced transparency in circuit quantum electrodynamics with nested polariton states. {\em Physical Review Letters}. \textbf{120}, 083602 (2018)
\bibitem{ann2020tunable}Ann, B. \& Steele, G. Tunable and weakly invasive probing of a superconducting resonator based on electromagnetically induced transparency. {\em Physical Review A}. \textbf{102}, 053721 (2020)
\bibitem{vadiraj2021engineering}Vadiraj, A., Ask, A., McConkey, T., Nsanzineza, I., Chang, C., Kockum, A. \& Wilson, C. Engineering the level structure of a giant artificial atom in waveguide quantum electrodynamics. {\em Physical Review A}. \textbf{103}, 023710 (2021)
\bibitem{chu2023three}Chu, K., Liao, W. \& Chen, Y. Three-level $\Lambda$-type microwave memory via parametric-modulation-induced transparency in a superconducting quantum circuit. {\em Physical Review Research}. \textbf{5}, 033192 (2023)
\bibitem{brehm2022slowing}Brehm, J., Gebauer, R., Stehli, A., Poddubny, A., Sander, O., Rotzinger, H. \& Ustinov, A. Slowing down light in a qubit metamaterial. {\em Applied Physics Letters}. \textbf{121} (2022)


\bibitem{roy2017colloquium}Roy, D., Wilson, C. \& Firstenberg, O. Colloquium: Strongly interacting photons in one-dimensional continuum. {\em Reviews Of Modern Physics}. \textbf{89}, 021001 (2017)
\bibitem{sheremet2023waveguide}Sheremet, A., Petrov, M., Iorsh, I., Poshakinskiy, A. \& Poddubny, A. Waveguide quantum electrodynamics: collective radiance and photon-photon correlations. {\em Reviews Of Modern Physics}. \textbf{95}, 015002 (2023)

\bibitem{kannan2023demand}Kannan, B., Almanakly, A., Sung, Y., Di Paolo, A., Rower, D., Braumüller, J., Melville, A., Niedzielski, B., Karamlou, A., Serniak, K. \& Others On-demand directional microwave photon emission using waveguide quantum electrodynamics. {\em Nature Physics}. \textbf{19}, 394-400 (2023)
\bibitem{grimsmo2021quantum}Grimsmo, A., Royer, B., Kreikebaum, J., Ye, Y., O’Brien, K., Siddiqi, I. \& Blais, A. Quantum metamaterial for broadband detection of single microwave photons. {\em Physical Review Applied}. \textbf{15}, 034074 (2021)








\bibitem{blais2007quantum}Blais, A., Gambetta, J., Wallraff, A., Schuster, D., Girvin, S., Devoret, M. \& Schoelkopf, R. Quantum-information processing with circuit quantum electrodynamics. {\em Physical Review A}. \textbf{75}, 032329 (2007)
\bibitem{strand2013first}Strand, J., Ware, M., Beaudoin, F., Ohki, T., Johnson, B., Blais, A. \& Plourde, B. First-order sideband transitions with flux-driven asymmetric transmon qubits. {\em Physical Review B}. \textbf{87}, 220505 (2013)
\bibitem{lu2017universal}Lu, Y., Chakram, S., Leung, N., Earnest, N., Naik, R., Huang, Z., Groszkowski, P., Kapit, E., Koch, J. \& Schuster, D. Universal stabilization of a parametrically coupled qubit. {\em Physical Review Letters}. \textbf{119}, 150502 (2017)
\bibitem{naik2017random}Naik, R., Leung, N., Chakram, S., Groszkowski, P., Lu, Y., Earnest, N., McKay, D., Koch, J. \& Schuster, D. Random access quantum information processors using multimode circuit quantum electrodynamics. {\em Nature Communications}. \textbf{8}, 1904 (2017)
\bibitem{chen2005manipulating}Chen, Y., Wang, S., Wang, C. \& Ite, A. Manipulating the retrieved width of stored light pulses. {\em Physical Review A}. \textbf{72}, 053803 (2005)

\bibitem{mucke2010electromagnetically}Mücke, M., Figueroa, E., Bochmann, J., Hahn, C., Murr, K., Ritter, S., Villas-Boas, C. \& Rempe, G. Electromagnetically induced transparency with single atoms in a cavity. {\em Nature}. \textbf{465}, 755-758 (2010)
\bibitem{specht2011single}Specht, H., Nölleke, C., Reiserer, A., Uphoff, M., Figueroa, E., Ritter, S. \& Rempe, G. A single-atom quantum memory. {\em Nature}. \textbf{473}, 190-193 (2011)
\bibitem{korber2018decoherence}Körber, M., Morin, O., Langenfeld, S., Neuzner, A., Ritter, S. \& Rempe, G. Decoherence-protected memory for a single-photon qubit. {\em Nature Photonics}. \textbf{12}, 18-21 (2018)







\bibitem{barends2013coherent}Barends, R., Kelly, J., Megrant, A., Sank, D., Jeffrey, E., Chen, Y., Yin, Y., Chiaro, B., Mutus, J., Neill, C. \& Others Coherent Josephson qubit suitable for scalable quantum integrated circuits. {\em Physical Review Letters}. \textbf{111}, 080502 (2013)


\bibitem{astafiev2010resonance}Astafiev, O., Zagoskin, A., Abdumalikov Jr, A., Pashkin, Y., Yamamoto, T., Inomata, K., Nakamura, Y. \& Tsai, J. Resonance fluorescence of a single artificial atom. {\em Science}. \textbf{327}, 840-843 (2010)






\bibitem{probst2015efficient}Probst, S., Song, F., Bushev, P., Ustinov, A. \& Weides, M. Efficient and robust analysis of complex scattering data under noise in microwave resonators. {\em Review Of Scientific Instruments}. \textbf{86} (2015)
\bibitem{lu2021characterizing}Lu, Y., Bengtsson, A., Burnett, J., Wiegand, E., Suri, B., Krantz, P., Roudsari, A., Kockum, A., Gasparinetti, S., Johansson, G. \& Others Characterizing decoherence rates of a superconducting qubit by direct microwave scattering. {\em Npj Quantum Information}. \textbf{7}, 35 (2021)
\bibitem{houck2007generating}Houck, A., Schuster, D., Gambetta, J., Schreier, J., Johnson, B., Chow, J., Frunzio, L., Majer, J., Devoret, M., Girvin, S. \& Others Generating single microwave photons in a circuit. {\em Nature}. \textbf{449}, 328-331 (2007)
\bibitem{blais2021circuit}Blais, A., Grimsmo, A., Girvin, S. \& Wallraff, A. Circuit quantum electrodynamics. {\em Reviews Of Modern Physics}. \textbf{93}, 025005 (2021)
\bibitem{johansson2012qutip}Johansson, J., Nation, P. \& Nori, F. QuTiP: An open-source Python framework for the dynamics of open quantum systems. {\em Computer Physics Communications}. \textbf{183}, 1760-1772 (2012)
\bibitem{li2013motional}Li, J., Silveri, M., Kumar, K., Pirkkalainen, J., Vepsäläinen, A., Chien, W., Tuorila, J., Sillanpää, M., Hakonen, P., Thuneberg, E. \& Others Motional averaging in a superconducting qubit. {\em Nature Communications}. \textbf{4}, 1-6 (2013)
\bibitem{beaudoin2012first}Beaudoin, F., Silva, M., Dutton, Z. \& Blais, A. First-order sidebands in circuit QED using qubit frequency modulation. {\em Physical Review A}. \textbf{86}, 022305 (2012)
\bibitem{gu2017microwave}Gu, X., Kockum, A., Miranowicz, A., Liu, Y. \& Nori, F. Microwave photonics with superconducting quantum circuits. {\em Physics Reports}. \textbf{718} pp. 1-102 (2017)
\bibitem{reim2010towards}Reim, K., Nunn, J., Lorenz, V., Sussman, B., Lee, K., Langford, N., Jaksch, D. \& Walmsley, I. Towards high-speed optical quantum memories. {\em Nature Photonics}. \textbf{4}, 218-221 (2010)
\bibitem{saglamyurek2018coherent}Saglamyurek, E., Hrushevskyi, T., Rastogi, A., Heshami, K. \& LeBlanc, L. Coherent storage and manipulation of broadband photons via dynamically controlled Autler–Townes splitting. {\em Nature Photonics}. \textbf{12}, 774-782 (2018)
\textbf{103}, 023710 (2021)
\bibitem{kervinen2019landau}
Kervinen, M., Ramírez-Muñoz, J., Välimaa, A., \& Sillanpää, M. (2019).
Landau-Zener-Stückelberg interference in a multimode electromechanical system in the quantum regime.
\emph{Physical Review Letters}, \textbf{123}, 240401.


\end{thebibliography}
\end{document}